# The work function of elastic solids

J. Garai and A. Laugier

**Abstract**  In solid phase the pressure correlates to the elastic related volume change while the temperature to the thermal related volume change.  These volume changes are not compatible with the exception of constant volume condition when the expanded volume converts completely compressed volume.  Separating the thermal and elastic related volume changes the work functions for each of the thermodynamic conditions are derived.  Based on theoretical consideration it is suggested that the thermal related volume change do not result mechanical work.  Homogeneous model, both the system and the surrounding have the same phase, can completely explain the lack of the thermal related work and provide a self-consistent thermodynamic description for the elastic solids.

J. Garai (✉) Department of Earth Sciences, University Park, PC 344, Miami, FL 33199, USA
E-mail: jozsef.garai@fiu.edu
A. Laugier · IdPCES - Centre d'affaires PATTON, 6, rue Franz Heller, bât. A, F35700 Rennes, France

---

## 1  Introduction

Historically the thermodynamic behavior of gas phase had been described first.  The thermodynamic state of an ideal gas phase can be completely described by the equation of state (EOS)

$$nRT = pV \qquad (1)$$

and by the first law of thermodynamics.

$$dU = \delta q + \delta w = nc_V dT - pdV \qquad (2)$$

In these equations n represents the number of moles, T is the absolute temperature, p is the pressure, V is the volume, R is the universal gas constant, U is the internal energy, q is the thermal energy, w is the mechanical energy, and $c_V$ is the molar heat capacity at constant volume. The phase transformation changes the physical properties of the substance. The thermodynamic relationships derived for gas phase have to be modified for solids.

Contrarily to gas phase no direct relationship between the pressure and the temperature is known in solid phase. The relationship between the temperature and the volume is characterized by the volume coefficient of expansion [$\alpha_{V_p}$]:

$$\alpha_{V_p} = \frac{1}{V}\left(\frac{\partial V}{\partial T}\right)_p \tag{3}$$

while the correlation between the pressure and the volume is described by the isothermal bulk modulus [$K_T$]:

$$K_T = -V\left(\frac{\partial p}{\partial V}\right)_T \tag{4}$$

It is assumed that the solid is homogeneous, isotropic, non-viscous and that the elasticity is linear. It is also assumed that the stresses are isotropic; therefore, the principal stresses can be identified as the pressure $p = \sigma_1 = \sigma_2 = \sigma_3$. In this text, the word solid will be used for materials complying with these criteria. In all of the calculations it will be assumed that the volume coefficient of expansion and the bulk modulus are constant.

$$\left(\frac{\partial K_T}{\partial V}\right)_T \approx 0 \; ; \; \left(\frac{\partial K_T}{\partial V}\right)_p \approx 0 \; ; \; \left(\frac{\partial K_T}{\partial p}\right)_V \approx 0 \; ; \; \left(\frac{\partial K_T}{\partial p}\right)_T \approx 0$$

$$\text{and} \quad \left(\frac{\partial \alpha_{V_p}}{\partial T}\right)_p \approx 0 \; ; \; \left(\frac{\partial \alpha_{V_p}}{\partial T}\right)_V \approx 0 \tag{5}$$

If this assumption can not be satisfied then first and higher order derivates of the volume coefficient of expansion and the bulk modulus has to be taken into consideration.

Contemporary thermodynamics assumes that beside the volume coefficient of expansion and the bulk modulus no additional adjustment to the gas equations is required and that the expression of the first law of thermodynamics can be used in the same form for solids as for gasses. It has been shown that using the contemporary description of elastic solids results path dependency for the internal energy of a system [1]. The sate function character of the internal energy cannot be challenged indicating that the current thermodynamic description of elastic solids is incomplete and/or incorrect. In this study the work function part of equation (2) will be considered in detail for solid phase. The thermal energy part of the equation will be discussed in a separate paper which will include the heat capacities and all the state functions.

Employing the equations of the volume coefficient of expansion and the bulk modulus the actual or total volume of a system can be calculated in two steps by allowing one of the variables to change while the other one held constant (Fig. 1).

$$(V_T)_{p=0} = V_0 e^{\int_{T=0}^{T} \alpha_{V_{p=0}} dT} \quad \text{or} \quad (V_p)_{T=0} = V_0 e^{\int_{p=0}^{p} -\frac{1}{K_{T=0}} dp} \qquad (6)$$

and then

$$(V_T)_p = (V_p)_{T=0} e^{\int_{T=0}^{T} \alpha_{V_p} dT} \quad \text{or} \quad (V_p)_T = (V_T)_{p=0} e^{\int_{p=0}^{p} -\frac{1}{K_T} dp} \qquad (7)$$

where $V_0$ is the volume at zero pressure and temperature. These two steps might be combined into one and the volume at a given p, and T can be calculated:

$$V_{p,T} = V_0 e^{\int_{T=0}^{T} \alpha_{V_p} dT - \int_{p=0}^{p} \frac{1}{K_T} dp} \qquad (8)$$

It has been shown that neither the pressure nor the temperature correlates to the actual volume [2] but rather to the elastic and thermal related volume changes respectively.

$$p = f(\Delta V_p^{elastic})_{(T=0)} \quad \text{and} \quad T = f(\Delta V_T^{thermal})_{(p=0)}. \tag{9}$$

The actual volume comprises the initial volume $(V_0)$, the elastic and the thermal related volume changes.

$$V = V_{p,T} = V_0 + \Delta V_{p,T}^{elastic} + \Delta V_{p,T}^{thermal}. \tag{10}$$

The elastic and thermal related volume changes are independent of each other since either of these volumes can be held constant while the other can still change.

$$(\Delta V_p^{elastic})_{(T=0)} \neq f_{link}\left[(\Delta V_T^{thermal})_{(p=0)}\right]. \tag{11}$$

The non-interchangeable nature of the thermal and elastic related volume results that the work function [Eq. (2)] must be modified and written in a more rigorous way.

$$\delta w(s)^{elastic} = -pdV^{elastic} \quad \text{and} \quad \delta w(s)^{thermal} = -pdV^{thermal} \tag{12}$$

where (s) is used for solid. The differentials of the volumes can be calculated from equations (3) and (4).

$$\left(dV_{p/V}^{elastic}\right)_T = -\frac{(V)_T}{K_T}dp \quad \text{and} \quad \left(dV_T^{thermal}\right)_p = \alpha_{V_p}(V)_p dT \tag{13}$$

The differential of the work is then the sum of the differentials of the elastic and the thermal related work.

$$\delta w(s) = \delta w(s)^{thermal} + \delta w(s)^{elastic} \tag{14}$$

The system is capable to do mechanical work under different conditions; therefore, the differential of the work should be written as :

$$\delta w(s) = \sum \delta w(s)^{thermal} + \sum \delta w(s)^{elastic} \tag{15}$$

The thermal and elastic work functions for each of the thermodynamic conditions will be derived below.

---

## 2  Thermal work

If the temperature is constant then thermal related volume change is zero. The zero thermal related volume change results no work.

$$\left(dV_{p/V}^{thermal}\right)_T \equiv 0 \quad \text{and} \quad \left[\delta w(s)_{p/V}^{thermal}\right]_T = -p\left(dV_{p/V}^{thermal}\right)_T = 0 \tag{16}$$

At zero pressure no mechanical work can be produced. Increasing the temperature at zero pressure will not result mechanical work either.

$$\left[\delta w(s)_T^{thermal}\right]_{p=0} = -p\left(dV_T^{thermal}\right)_{p=0} = 0 \tag{17}$$

Any thermodynamic conditions can be reached by increasing the temperature at zero pressure and then applying pressure at constant temperature.

$$\text{step 1} \left[(T=0 \Rightarrow T=T)_{p=0}\right] \quad \text{and} \quad \text{step 2} \left[(p=0 \Rightarrow p=p)_{T=\text{constant}}\right] \tag{18}$$

Since neither of these thermodynamic processes produces mechanical work the thermal related work should remain zero between any conditions. Despite this conclusion substituting equation (13) into equation (12) indicates the existence of thermal work.

$$\left[\delta w(s)_T^{thermal}\right]_p = -p\left(dV_T^{thermal}\right)_p = -p\alpha_{V_p} V dT \tag{19}$$

Factorizing the volume and integrating the equation gives the work done by/on the system between the initial and final conditions.

$$\left[\Delta w(s)_T^{thermal}\right]_p = \int_{T_i}^{T_f} -p\alpha_{V_p} V_0 e^{\int_{p=0}^{p} -\frac{1}{K_T} dp} e^{\int_{T=0}^{T} \alpha_{V_p} dT} dT = -p\alpha_{V_p} V_0 e^{\int_{p=0}^{p} -\frac{1}{K_T} dp} \int_{T_i}^{T_f} e^{\alpha_{V_p} T} dT \tag{20}$$

$$\Rightarrow \left[\Delta w(s)_T^{\text{thermal}}\right]_p = -p\alpha_{V_p} V_0 e^{\int_{p=0}^{p} -\frac{1}{K_T} dp} \left[\frac{1}{\alpha_{V_p}} e^{\alpha_{V_p} T}\right]_{T_i}^{T_f} = -pV_0 e^{\int_{p=0}^{p} -\frac{1}{K_T} dp} \left[e^{\alpha_{V_p} T_f} - e^{\alpha_{V_p} T_i}\right]$$

$$\Rightarrow \left[\Delta w(s)_T^{\text{thermal}}\right]_p = -pV_0 e^{\int_{p=0}^{p} -\frac{1}{K_T} dp} e^{\alpha_{V_p} T_i} \left[e^{\alpha_{V_p} (T_f - T_i)} - 1\right]$$

In equation (20) the non zero thermal related work is in discrepancy with equations (16)-(17). If equation (20) is correct then the internal energy of the system becomes path dependent[1]. The path dependency of the internal energy would violate the conservation law of energy. The conservation law of energy is one of the fundamental principles of physics and has never been shown to be inexact. The contradiction with this basic principle indicates that equation (20) must be incorrect.

The lack of thermal related work can be explained by a homogenous phase model. In this model the system and the surrounding have the same phase. Constant pressure or volume in this solid-solid system can be maintained by applying constrains on the temperature of the surrounding. Maintaining the same temperature between the system and the surrounding will result constant pressure. The identical temperature between the system and surrounding results synchronized expansion with no thermal related mechanical work when temperature is changed. The constant volume of the system can be maintained by fixing the temperature of the surrounding. If the temperature remains the same then the volume of the surrounding would not change. Ideal conditions are assumed and the developing additional pressure in the system should not deform the surrounding.

## 3 Elastic work

### 3.1 Constant temperature

It has been shown that the thermal related work is always zero. The differential of the work [Eq. (12)-(15)] contains only the elastic work; therefore, the superscript elastic will be omitted.

$$\delta w(s) = \delta w(s)^{elastic} \tag{21}$$

The isothermal work change can be calculated as:

$$[\delta w(s)]_T = -p(dV_{p/V}^{elastic})_T \tag{22}$$

Substituting equation (13) into equation (12) gives:

$$[\delta w(s)]_T = \frac{(V_p)_T}{K_T} p\, dp. \tag{23}$$

Using equation (6)-(8) $(V_p)_T$ can be written as:

$$(V_p)_T = (V_T)_{p=0}\, e^{\int_{T=0}^{T} \alpha_{V_p} dT} = V_0\, e^{\int_{T=0}^{T} \alpha_{V_p} dT}\, e^{\int_{p=0}^{p} -\frac{1}{K_T} dp} \tag{24}$$

Substituting the factorized volume expression into equation (23) and integrating between the initial and final conditions the work done by or on the system between these states can be determined.

$$[\Delta w(s)_p]_T = \frac{V_0\, e^{\int_{T=0}^{T} \alpha_{V_p} dT}}{K_T} \int_{p_i}^{p_f} p\, e^{\int_{p=0}^{p} -\frac{1}{K_T} dp}\, dp \tag{25}$$

The work between the initial and final conditions is [3]:

$$[\Delta w(s)_p]_T = \frac{V_0\, e^{\int_{T=0}^{T} \alpha_{V_p} dT}}{K_T} \left\{ -K_T \left[ p\, e^{-\frac{p}{K_T}} \right]_{p_i}^{p_f} + K_T \int_{p_i}^{p_f} e^{-\frac{p}{K_T}}\, dp \right\} \tag{26}$$

$$= \frac{V_0 e^{\int_{T=0}^{T} \alpha_{V_p} dT}}{K_T} \left\{ -K_T \left[ p_i e^{-\frac{p_i}{K_T}} - p_f e^{-\frac{p_f}{K_T}} \right] + K_T^2 \left[ p e^{-\frac{p}{K_T}} \right]_{p_i}^{p_f} \right\}$$

$$= V_0 e^{\int_{T=0}^{T} \alpha_{V_p} dT} \left[ p_i e^{-\frac{p_i}{K_T}} - p_f e^{-\frac{p_f}{K_T}} - K_T e^{-\frac{p_f}{K_T}} + K_T e^{-\frac{p_i}{K_T}} \right].$$

Rearranging the terms in (26), it gives:

$$\left[ \Delta w(s)_p \right]_T = V_0 e^{\int_{T=0}^{T} \alpha_{V_p} dT} \left[ (K_T + p_i) e^{-\frac{p_i}{K_T}} - (K_T + p_f) e^{-\frac{p_f}{K_T}} \right] \qquad (27)$$

Using [Eq. (6)-(8)] the initial and final volumes are:

$$V_i = V_0 e^{\int_{T=0}^{T} \alpha_{V_p} dT} e^{\int_{p=0}^{p_i} -\frac{1}{K_T} dp} \quad \text{and} \quad V_f = V_0 e^{\int_{T=0}^{T} \alpha_{V_p} dT} e^{\int_{p=0}^{p_f} -\frac{1}{K_T} dp} \qquad (28)$$

Then substituting these volumes into (27), it gives:

$$\left[ \Delta w(s)_{p/V} \right]_T = V_i (K_T + p_i) - V_f (K_T + p_f) \qquad (29)$$

The function between the pressure and the volume at constant temperature allows determining the work from either the volume changes or the pressure changes.

$$V \equiv (V_p)_T = (V_{p=0})_T e^{-\frac{p}{K_T}} \iff p = -K_T \ln\left[\frac{(V)_T}{(V_{p=0})_T}\right] \qquad (30)$$

Using the volume as a variable the infinitesimal isothermal solid work change in [Eq. (12)] becomes:

$$\left[ \delta w(s)_V \right]_T = K_T \ln\left[\frac{(V)_T}{(V_{p=0})_T}\right] (dV^{elastic})_T \qquad (31)$$

The work between the initial and final conditions can be determined by integrating equation (31).

$$[\Delta w(s)_V]_T = K_T \int_{V_i}^{V_f} \ln\left[\frac{(V)_T}{(V_{p=0})_T}\right](dV^{elastic})_T = K_T(V_{p=0})_T \int_{x_i}^{x_f} \ln x \, dx \qquad (32)$$

where the dimensionless variable x is defined as:

$$x = \frac{(V)_T}{(V_{p=0})_T} \qquad (33)$$

Performing the integration [Eq. (32)], it gives:

$$[\Delta w(s)_V]_T = K_T(V_{p=0})_T [x \ln x - x]_{x_i}^{x_f} = K_T\left[V_f \ln \frac{V_f}{(V_{p=0})_T} - V_i \ln \frac{V_i}{(V_{p=0})_T} + V_i - V_f\right] \qquad (34)$$

Using (28) and substituting into (34), it gives:

$$[\Delta w(s)_{p/V}]_T = K_T\left[-\frac{p_f V_f}{K_T} + \frac{p_i V_i}{K_T} + V_i - V_f\right] \qquad (35)$$

$$\Rightarrow [\Delta w(s)_{p/V}]_T = V_i(K_T + p_i) - V_f(K_T + p_f)$$

This is exactly the equation (29). The equivalency between equation (29) and (35) is proof of that the function between the pressure and compressed volume allows using explicitly either the pressure or the compressed volume to determine the mechanical work. The total elastic work of the system at constant temperature can be calculated from the volume by using equation (34)

$$[\Delta w(s)_V^{total}]_T = K_T\left\{V\left[\ln \frac{V}{(V_{p=0})_T} - 1\right] + (V_{p=0})_T\right\} \qquad (36)$$

or from the pressure by using equation (27)

$$[\Delta w(s)_p^{total}]_T = V_0 e^{\int_{T=0}^{T} \alpha_{V_p} dT}\left[K_T - (K_T + p)e^{-\frac{p}{K_T}}\right] \qquad (37)$$

In equations (36) and (37) it was assumed that

$$V_i = (V_{p=0})_T, \quad V = V_f \quad \text{and} \quad p_i = 0, \; p_f = p \tag{38}$$

3.2. Constant pressure

When the volume expands at constant pressure the expanded part of the volume is compressed by the existing pressure. This compressed part of the expanded volume (Fig. 1) results mechanical work, which will be approximated as:

$$\left[\delta w(s)_T\right]_p \approx -\frac{1}{2}p\left(dV_T^{elastic}\right)_p \tag{39}$$

Integrating the work between the initial and final conditions gives the mechanical energy resulting from the expansion.

$$\left[\Delta w(s)_T\right]_p \approx \int_{T_i}^{T_f} -\frac{p}{2}\left(dV_T^{elastic}\right)_p = -\frac{p}{2}\left(\Delta V_{\Delta T}^{elastic}\right)_p. \tag{40}$$

where $\Delta T = T_f - T_i$. The expansion at zero pressure is

$$\left(\Delta V_{\Delta T}^{thermal}\right)_{p=0} = V_0 e^{\int_{T=0}^{T_i} \alpha_{V_p} dT}\left(e^{\int_{T_i}^{T_f} \alpha_{V_p} dT} - 1\right) \tag{41}$$

The compressed part of this expansion at pressure p is

$$\left(\Delta V_{\Delta T}^{elastic}\right)_p = \left(\Delta V_{\Delta T}^{thermal}\right)_{p=0}\left(e^{\int_{p=0}^{p} -\frac{1}{K_T} dp} - 1\right) = V_0 e^{\int_{T=0}^{T_i} \alpha_{V_p} dT}\left(e^{\int_{T_i}^{T_f} \alpha_{V_p} dT} - 1\right)\left(e^{\int_{p=0}^{p} -\frac{1}{K_T} dp} - 1\right) \tag{42}$$

The work resulting from this expansion is

$$\left[\Delta w(s)_T\right]_p \approx -\frac{pV_0}{2} e^{\int_{T=0}^{T_i} \alpha_{V_p} dT}\left(e^{\int_{T_i}^{T_f} \alpha_{V_p} dT} - 1\right)\left(e^{\int_{p=0}^{p} -\frac{1}{K_T} dp} - 1\right). \tag{43}$$

Calculating the work as the area of a triangle assumes that the pressure linearly correlates to the compressed volume. The bulk modulus defines the pressure volume relationship a slightly different way by assuming linearity between the pressure and the strain [Eq. (4)].

Using the relationship given by the bulk modulus the total work at a given pressure and temperature can be determined from equation (37). Calculating the total work for the same pressure at different temperatures and deducting one from the other gives the work done between the initial and final temperatures.

$$[\Delta w(s)_{\Delta T=T_f-T_i}]_p = [\Delta w(s)^{total}]_{p,T_f} - [\Delta w(s)^{total}]_{p,T_i} \tag{44}$$

Using the initial and final temperature in equation (44) gives the precise value of the work done at constant pressure.

$$[\Delta w(s)_{\Delta T}]_p = V_0 \left[ K_T - (K_T + p) \, e^{-\frac{p}{K_T}} \right] \left( e^{\int_{T=0}^{T_f} \alpha_{V_p} dT} - e^{\int_{T=0}^{T_i} \alpha_{V_p} dT} \right)$$

$$= V_0 e^{\int_{T=0}^{T_i} \alpha_{V_p} dT} \left[ K_T - (K_T + p) \, e^{-\frac{p}{K_T}} \right] \left( e^{\int_{T_i}^{T_f} \alpha_{V_p} dT} - 1 \right) \tag{45}$$

The difference between equation (43) and (45) are minor but the theoretically correct solution is given by equation (45).

3.3. Constant volume

At constant volume the volume increase caused by the thermal expansion has to be canceled by the volume decrease resulting from the compression.

$$V = V_0 e^{\int_{p=0}^{p_i} -\frac{1}{K_T} dp} e^{\int_{T=0}^{T_i} \alpha_{V_p} dT} = V_0 e^{\int_{p=0}^{p_f} -\frac{1}{K_T} dp} e^{\int_{T=0}^{T_f} \alpha_{V_p} dT} \tag{46}$$

The work will be calculated in two steps by using the previously derived work functions for constant pressure and for constant temperature. First at constant pressure the volume will be allowed to expand or contract while the temperature is increased or decreased respectively. The work done during this expansion or contraction part $[\Delta w(s)_T^{elastic}]_p$ can be calculated by using equation (45).

$$[\Delta w(s)_T]_p = V_0 \left[ K_T - (K_T + p)\, e^{-\frac{p}{K_T}} \right] \left( e^{\int_{T=0}^{T_f} \alpha_{V_p} dT} - e^{\int_{T=0}^{T_i} \alpha_{V_p} dT} \right) \qquad (47)$$

In step two the pressure will be changed until the volume reaches its original size. The expended or contracted volume can be calculated as:

$$\left(\Delta V_{T_i - T_f}\right)_{p_i} = V_0\, e^{\int_{p=0}^{p_i} -\frac{1}{K_T} dp}\, e^{\int_{T=0}^{T_i} \alpha_V dT} \left( e^{\int_{T_i}^{T_f} \alpha_V dT} - 1 \right) \qquad (48)$$

The initial volume $[(V_i)_T]$ at pressure $p_i$ is then

$$V_i = V + \left(\Delta V_{T_i - T_f}\right)_{p_i} = V_0\, e^{\int_{p=0}^{p_i} -\frac{1}{K_T} dp}\, e^{\int_{T=0}^{T_i} \alpha_V dT} + V_0\, e^{\int_{p=0}^{p_i} -\frac{1}{K_T} dp}\, e^{\int_{T=0}^{T_i} \alpha_V dT} \left( e^{\int_{T_i}^{T_f} \alpha_V dT} - 1 \right) \qquad (49)$$

$$V_i = V_0\, e^{\int_{p=0}^{p_i} -\frac{1}{K_T} dp}\, e^{\int_{T=0}^{T_f} \alpha_V dT}$$

while the final volume $[(V_f)_T]$ at pressure $p_f$ is equal with the original volume. Substituting the initial and final volumes into equation (36) and simplifying the equation gives the compression work:

$$\begin{aligned}[][\Delta w(s)_{p/V}]_T &= K_T V_0\, e^{\int_{p=0}^{p_i} -\frac{1}{K_T} dp}\, e^{\int_{T=0}^{T_i} \alpha_{V_p} dT} \left[ e^{\int_{T_i}^{T_f} \alpha_{V_p} dT} \left(1 + \frac{p_i}{K_T}\right) - \left(1 + \frac{p_f}{K_T}\right) \right] \\ &= V K_T \left[ e^{\int_{T_i}^{T_f} \alpha_{V_p} dT} \left(1 + \frac{p_i}{K_T}\right) - \left(1 + \frac{p_f}{K_T}\right) \right] \end{aligned} \qquad (50)$$

where the final pressure can be calculated as :

$$p_f = p_i - K_T \ln \frac{V}{V + \left(\Delta V_{T_i - T_f}\right)_{p_i}} = p_i - K_T \ln \frac{1}{e^{\int_{T_i}^{T_f} \alpha_{V_p} dT}} = p_i + K_T \int_{T_i}^{T_f} \alpha_{V_p} dT \qquad (51)$$

Substituting the final pressure into equation (50) gives :

$$\left[\Delta w(s)_{p/V}\right]_T = V_0 e^{-\frac{p_i}{K_T}} e^{\alpha_{V_p} T_i} K_T \left\{ e^{\int_{T_i}^{T_f} \alpha_{V_p} dT} \left[1 + \frac{p_i}{K_T}\right] - \left[1 + \frac{p_i + \alpha_{V_p} K_T (T_f - T_i)}{K_T}\right] \right\} \quad (52)$$

Adding the expansion and the compression part of the work gives the work done on or by the system at constant volume.

$$\left[\Delta w(s)_T\right]_V = V_0 e^{-\frac{p_i}{K_T}} e^{\alpha_{V_p} T_i} K_T \left\{ e^{\int_{T_i}^{T_f} \alpha_{V_p} dT} \left[1 + \frac{p_i}{K_T}\right] - \left[1 + \frac{p_i + \alpha_{V_p} K_T (T_f - T_i)}{K_T}\right] \right\}$$
$$+ V_0 \left[K_T - (K_T + p) e^{-\frac{p}{K_T}}\right] \left( e^{\int_{T=0}^{T_f} \alpha_{V_p} dT} - e^{\int_{T=0}^{T_i} \alpha_{V_p} dT} \right) \quad (53)$$

Rearranging and simplifying equation (53), it remains :

$$\left[\Delta w(s)_T\right]_V = K_T V_0 e^{\alpha_{V_p} T_i} \left[ e^{\int_{T_i}^{T_f} \alpha_{V_p} dT} - 1 - \alpha_{V_p} e^{-\frac{p_i}{K_T}} (T_f - T_i) \right] \quad (54)$$

---

## 4  Conclusions

Based on theoretical considerations it is suggested that in solid phase only the elastic related volume changes do work.  The existence of purely elastic work can be explained by a homogeneous solid-solid model.  The equations describing the elastic work of solid systems had been derived for each of the thermodynamic conditions.

**References**

1. J. Garai, and A. Laugier (2005) physics/0508107
2. J. Garai (2005) physics/0507075
3. The product of polynomial p(x) with $e^x$ can be integrated by using the product rules of the derivates.  The integration can be done by parts.
$$\int p(x) e^{ax} dx = \frac{1}{a} p(x) e^{ax} - \frac{1}{a} \int p'(x) e^{ax} dx = \frac{1}{a} p(x) e^{ax} - \frac{1}{a^2} p'(x) e^{ax} + \frac{1}{a^3} p''(x) e^{ax} - ...$$
where the signs alternate as: (+-+-...).

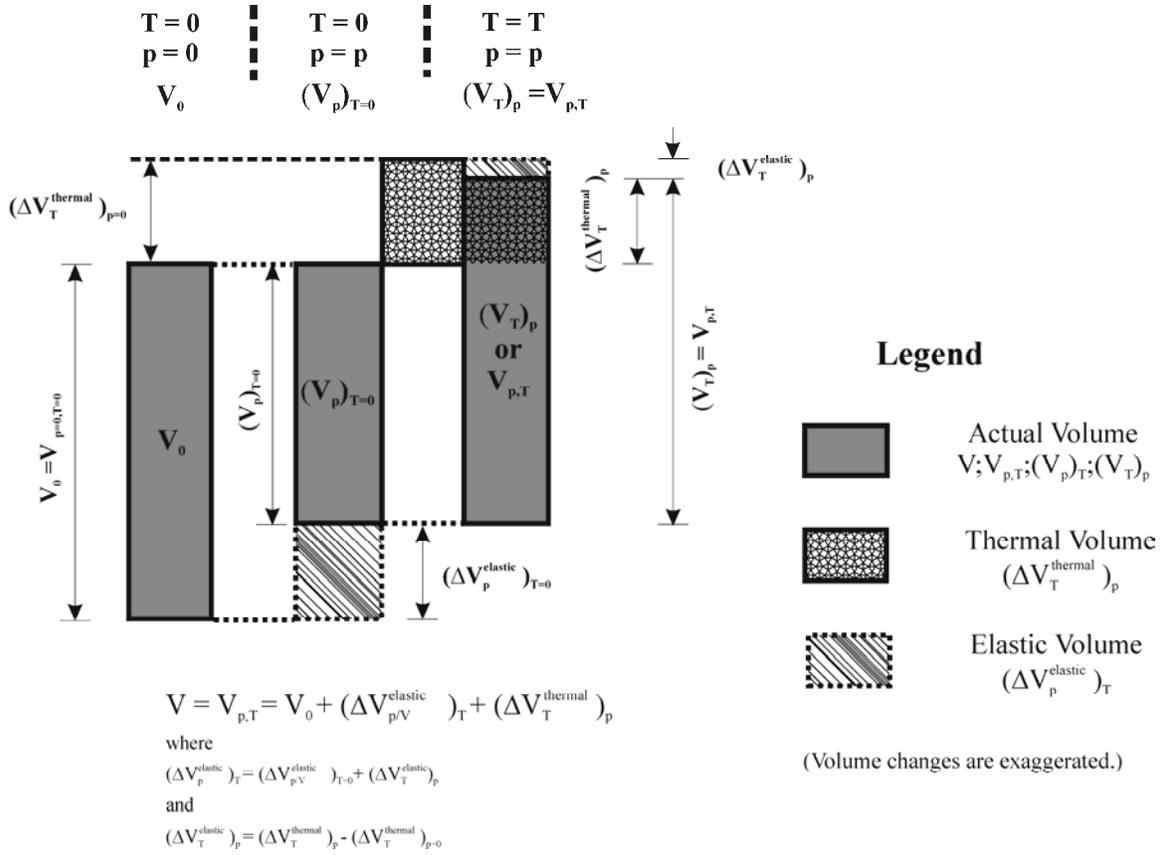

**Fig. 1.** Schematic figure of the volume changes caused by the pressure and the temperature. The analytical description of the different volumes is given in the text with the exceptions of the thermal volume at zero pressure $\left(\Delta V_T^{thermal}\right)_{p=0}$, the elastic volume at zero temperature $\left(\Delta V_p^{elastic}\right)_{T=0}$ and the elastic part of the thermal volume $\left(\Delta V_T^{elastic}\right)_p$, which are given here: $\left(\Delta V_p^{elastic}\right)_{T=0} = V_0\left(e^{\int_{p=0}^{p}\frac{dp}{K_T}} - 1\right)$; $\left(\Delta V_T^{thermal}\right)_{p=0} = V_0\left(e^{\int_{T=0}^{T}\alpha_V dT} - 1\right)$, and

$$\left(\Delta V_T^{elastic}\right)_p = V_0\left(e^{\int_{p=0}^{p}\frac{dp}{K_T}} - 1\right)\left(e^{\int_{T=0}^{T}\alpha_V dT} - 1\right).$$